\documentclass{article}
\usepackage{spconf,amsmath,graphicx}
\usepackage{hyperref}
\hypersetup{colorlinks=true}
\usepackage{multirow}
\usepackage{amssymb}
\usepackage{float}
\usepackage{array}
\usepackage{enumitem}
\usepackage{amsthm}
\usepackage[export]{adjustbox}
\usepackage[caption=false]{subfig}
\newcommand\mat[1]{\mathbf{#1}}
\def\L{{\cal L}}
\renewcommand{\paragraph}[1]{\noindent\textbf{\textit{#1}}\quad}

\title{Multi-user VoiceFilter-Lite via Attentive Speaker Embedding}
\name{Rajeev Rikhye$^{*}$, Quan Wang$^{*}$, Qiao Liang, Yanzhang He, Ian McGraw\thanks{* Equal contribution. }}
\address{
Google LLC, USA \\
\small \texttt{\{\href{mailto:rvrikhye@google.com}{rvrikhye},\href{mailto:quanw@google.com}{quanw}\}@google.com}
}

\begin{document}
\ninept
\maketitle
\begin{abstract}
In this paper, we propose a solution to allow speaker conditioned speech models, such as VoiceFilter-Lite, to support an arbitrary number of enrolled users in a single pass. This is achieved via an attention mechanism on multiple speaker embeddings to compute a single attentive embedding, which is then used as a side input to the model. We implemented multi-user VoiceFilter-Lite and evaluated it for three tasks: (1) a streaming automatic speech recognition (ASR) task; (2) a text-independent speaker verification task; and (3) a personalized keyphrase detection task, where ASR has to detect keyphrases from multiple enrolled users in a noisy environment. Our experiments show that, with up to four enrolled users, multi-user VoiceFilter-Lite is able to significantly reduce speech recognition and speaker verification errors when there is overlapping speech, without affecting performance under other acoustic conditions. This attentive speaker embedding approach can also be easily applied to other speaker-conditioned models such as personal voice activity detection (VAD) and personalized ASR. 
\end{abstract}
\section{Introduction}
\label{sec:intro}

Many speech models rely on a target speaker embedding as a side input to achieve either better performance or a more personalized user experience. This target speaker embedding is usually obtained via an offline enrollment process~\cite{enrollmentblog,wang2020version}, and is directly usable at runtime. We refer to this type of speech models as \emph{speaker-conditioned models}; in comparison, we refer to speech models that do not rely on target speaker embeddings as \emph{generic models}.

For example, speech enhancement or separation models can leverage the target speaker embedding to avoid the complexity of permutation invariant training~\cite{yu2017permutation}, and directly output the clean waveform of the target speaker. Representative works include DENet~\cite{wang2018deep}, SpeakerBeam ~\cite{zmolikova2017speaker,vzmolikova2017learning,delcroix2018single}, VoiceFilter~\cite{Wang2019} and more recently SpEx~\cite{xu2020spex}. 

VoiceFilter-Lite is a variant of speaker-conditioned speech separation models that is optimized for streaming on-device applications, such as automatic speech recognition (ASR)~\cite{Wang2020} and text-independent speaker verification (TI-SV)~\cite{rikhye2021personalized}. It directly takes the acoustic frontend features of ASR (\emph{e.g.} stacked log Mel-filterbank energies) as an input and outputs enhanced filterbank energies, instead of operating on the audio waveforms.

Target speaker embedding can also be used for voice activity detection (VAD) and speech recognition. In a personal VAD system~\cite{ding2019personal}, an always-on, small-footprint VAD model can be used to reject frames containing either non-speech noise or speech from a non-target speaker. In scenarios where a keyword detector is not preferable, personal VAD largely reduces false triggering rate of more expensive downstream models such as ASR and speaker recognition, thus saving computational cost and power consumption. In a target-speaker speech ASR system~\cite{denisov2019end,shi2021improving}, the target speaker embedding helps the ASR model to better recognize the speech from the target speaker under noisy environments with background noise and interfering speakers.

Most of the above mentioned works assume that only a single target speaker embedding is present at runtime. This assumption is usually valid for personal devices, such as mobile phones. However, for shared devices, such as smart home speakers and displays, multiple users can enroll their voices on the device, and thus more than one target speaker embeddings are usually present at runtime. In such cases, we need to either: (1) disable the personalized model and fallback to a generic model; or (2) run the entire system for multiple passes for each enrolled speaker embedding and choose the best output, which can be extremely expensive and unacceptable for on-device applications.

In this paper, we address the multi-user challenge for speaker-conditioned models by leveraging an attention mechanism~\cite{vaswani2017attention} on the enrolled speaker embeddings.
We implemented a multi-user VoiceFilter-Lite model that can significantly improve speech recognition and speaker verification performance when the input audio contains overlapped speech, and can take an arbitrary number of enrolled speaker embeddings as side input.
This multi-user VoiceFilter-Lite model is also critical to a personalized keyphrase detection system on shared devices. Apart from VoiceFilter-Lite, we would like to point out that the same mechanism can be easily applied to various other applications, such as waveform-based speech separation, personal VAD and personalized ASR.

The idea of ``attentive speaker embedding'' can also be found in other works, where the attention mechanism is mostly used as a pooling layer to aggregate frame-level embeddings in the \underline{same utterance}~\cite{wang2018attention,rahman2018attention,an2019deep,xu2021target}. To avoid confusion, we emphasize that in this work, the attentive speaker embedding attends to multiple embeddings from \underline{different speakers} that are already obtained via a separate offline enrollment process, which is more relevant to the work in~\cite{pan2018online,wan2020speaker}. Using an attention mechanism over an inventory of speaker embeddings was proposed in two recent papers~\cite{kanda2020joint,wang2020speaker}, where the target speaker profile(s) are selected by maximizing a similarity coefficient (correlation~\cite{wang2020speaker} and cosine similarity~\cite{kanda2020joint}) between each speaker embedding in the inventory and an embedding computed from overlapping speech. Unlike these papers, we use a \emph{ScorerNet} to compute the posterior probability of the target speaker speaking in a given frame conditioned on all other speakers, and our method is non-iterative allowing it to be used in a streaming fashion.

\section{Methods}

% \begin{figure*}[t!]
%     \subfloat[\label{fig:single_embedding}]{%
%       \includegraphics[width=0.48\textwidth]{single_embedding.pdf}
%      }
%      \subfloat[\label{fig:attentive_embedding}]{%
%       \includegraphics[width=0.51\textwidth]{attentive_embedding.pdf}
%      }
%      \caption{Attentive speaker embedding in VoiceFilter-Lite. (a) Original VoiceFilter-Lite, where the d-vector of the single target speaker is frame-wise concatenated to the input features frame}
%      \label{fig:embeddings}
% \end{figure*}

\begin{figure}[t!]
    \subfloat[\label{fig:single_embedding}]{%
      \includegraphics[width=0.49\textwidth]{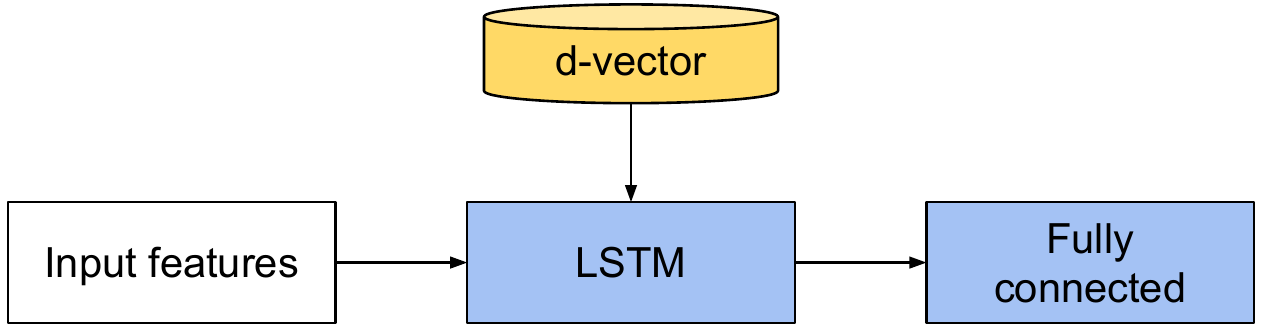}
     }\\
     \subfloat[\label{fig:attentive_embedding}]{%
      \includegraphics[width=0.49\textwidth]{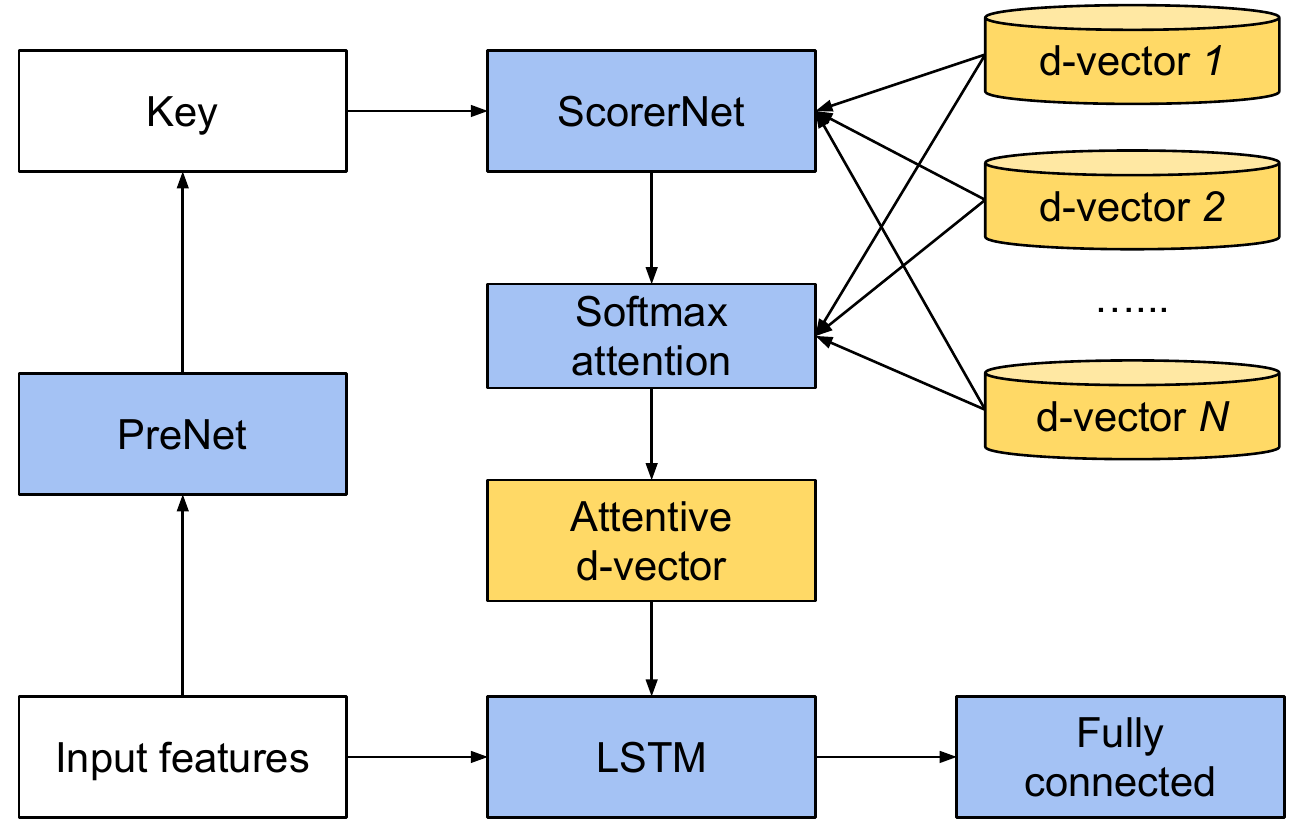}
     }
     \caption{Attentive speaker embedding in VoiceFilter-Lite. (a) Original VoiceFilter-Lite, where the d-vector of the single target speaker is frame-wise concatenated to the input features; (b) Multi-user VoiceFilter-Lite, where we support an arbitrary number $N$ of d-vectors (the target speaker embedding is included in these $N$ d-vectors).}
     \label{fig:embeddings}
\end{figure}

\subsection{Review of VoiceFilter-Lite}
VoiceFilter-Lite is a targeted voice separation model for streaming on-device speech recognition~\cite{Wang2020}, as well as text-independent speaker verification (TI-SV)~\cite{rikhye2021personalized}. It assumes that the target speaker has completed an offline enrollment process, which uses a speaker recognition model (see Section~\ref{sec:spk_reg}) to produce an aggregated embedding vector $\mat{e}$ that presents the voice characteristics of this speaker. In this work, we use the d-vector embedding~\cite{wan2018generalized} trained with the generalized end-to-end extended-set softmax loss~\cite{pelecanos2021dr}.

Let $\mat{x}^{(t)}$ be the input feature frame at time $t$. This feature is first frame-wise concatenated with the d-vector $\mat{e}$, then fed into an LSTM network~\cite{hochreiter1997long} followed by a fully connected neural network to produce a mask $\mat{y}^{(t)}$ (see Fig.~\ref{fig:single_embedding}):
\begin{equation}
    \label{eq:voicefilterlite}
    \mat{y}^{(t)} = \mathrm{FC} \circ \mathrm{LSTM} ( \mathrm{Concat} (\mat{x}^{(t)}, \mat{e}) ) .
\end{equation}

At runtime, the mask $\mat{y}^{(t)}$ is element-wise multiplied to the input $\mat{x}^{(t)}$ to produce the final enhanced features. Separately, we also use another LSTM-based neural network followed by a fully connected layer to estimate the noise type (either overlapping or non-overlapping speech) from the input $\mat{x}^{(t)}$. This noise type prediction is then used during inference to deactivate the VoiceFilter model when the input frame contains no overlapping speech. For more details, we refer the reader to~\cite{Wang2020}.

\subsection{Attentive embedding selection}
In the multi-user scenario, we have multiple enrolled speaker embeddings $\mat{e}_1, \mat{e}_2, \cdots, \mat{e}_N$. First, we use another small LSTM-based neural network which we refer to as the \emph{PreNet} to compute a \emph{key vector} $\mat{k}^{(t)}$ for each frame:
\begin{equation}
    \label{eq:prenet}
    \mat{k}^{(t)} = \mathrm{PreNet}(\mat{x}^{(t)}) .
\end{equation}
We concatenate this key vector with each speaker embedding, and feed it as an input to a fully connected neural network which we refer to as the \emph{ScorerNet} to produce attention scores $s_i^{(t)} \in \mathbb{R}$:
\begin{equation}
    \label{eq:scorer_net}
    s_i^{(t)} = \mathrm{ScorerNet}(\mathrm{Concat}(\mat{k}^{(t)}, \mat{e}_i)) .
\end{equation}
We use a softmax layer to convert these scores into attention weights $\alpha_i^{(t)} > 0$:
\begin{equation}
    \label{eq:softmax}
    \alpha_i^{(t)} = \frac{\exp{(s_i^{(t)})}}{\sum _{j=1}^N \exp{(s_j^{(t)})}} .
\end{equation}
And finally, the attentive speaker embedding $\mat{e}_\mathrm{att}^{(t)}$ at time $t$ is the weighted sum of all enrolled speaker embeddings:
\begin{equation}
    \label{eq:weighted_average}
    \mat{e}_\mathrm{att}^{(t)} = \sum_{i=1}^N \alpha_i^{(t)} \cdot \mat{e}_i .
\end{equation}

With the attentive speaker embedding, the rest of multi-user VoiceFilter-Lite is similar to the original VoiceFilter-Lite as described in Eq.~\ref{eq:voicefilterlite} (see Fig.~\ref{fig:attentive_embedding}):
\begin{equation}
    \label{eq:muvf}
    \mat{y}^{(t)} = \mathrm{FC} \circ \mathrm{LSTM} ( \mathrm{Concat} (\mat{x}^{(t)}, \mat{e}_\mathrm{att}^{(t)}) ) .
\end{equation}

\subsection{Loss function}

The original VoiceFilter-Lite model is trained with two losses~\cite{Wang2020}:
\begin{enumerate}
    \item $L_\mathrm{asym}$: an asymmetric L2 loss  for signal reconstruction;
    \item $L_\mathrm{noise}$: a noise type prediction loss  for adaptive suppression at runtime.
\end{enumerate}

For multi-user VoiceFilter-Lite, we introduce a new loss that measures how well the attentive embedding selection module attends to the correct target embedding:
\begin{equation}
    \label{eq:att_loss}
    L_\mathrm{att}= \sum _ t || \mat{e}_\mathrm{att}^{(t)} - \mat{e}_\mathrm{tar} ||^2, 
\end{equation}
where $\mat{e}_\mathrm{tar} \in \{\mat{e}_1, \cdots, \mat{e}_N\}$ is the target speaker embedding. The final loss for multi-user VoiceFilter-Lite training is then a linear combination of $L_\mathrm{asym}$, $L_\mathrm{noise}$, and $L_\mathrm{att}$. The weights for this combination were determined by tuning performance on a validation set. We use the Adam optimizer~\cite{kingma2014adam} to optimize our loss function.

\subsection{Design philosophy}

\begin{figure*}
	\centering
	\includegraphics[width=0.9\textwidth]{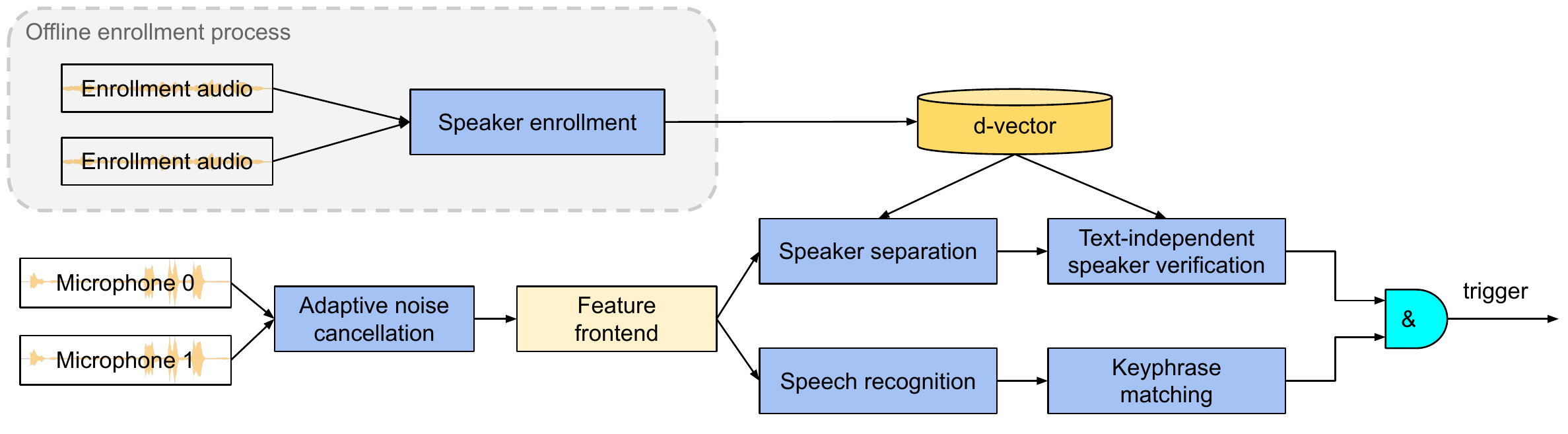}
	\caption{Diagram of the personalized keyphrase detection system. The d-vector is obtained in a separate offline enrollment process.}
	\label{fig:digram}
\end{figure*}

\subsubsection{PreNet}
The purpose of the PreNet is to produce a key vector that represents the voice characteristics of the input features, similar to a speaker recognition neural network. Since both the PreNet and the main VoiceFilter-Lite network are based on low-level input features, it's possible to share common low-level layers between these two networks, such as 1-D convolutional layers.

\subsubsection{Permutation invariance}

The softmax attention mechanism in Eq.~\ref{eq:softmax} and Eq.~\ref{eq:weighted_average} guarantees that the inference and training process of the model is invariant to permutations of the speaker embeddings. This is critical for efficient training.

An alternative solution  we also considered is to directly concatenate all the enrolled speaker embeddings into a single super embedding:
\begin{equation}
    \label{eq:super_embedding}
    \mat{e}_\mathrm{sup} = \mathrm{Concat}(\mat{e}_1, \mat{e}_2, \cdots, \mat{e}_N) ,
\end{equation}
and replace the embedding in Eq.~\ref{eq:voicefilterlite} by this super embedding instead of an attentive embedding. This solution has several drawbacks compared with the attentive embedding solution:
\begin{enumerate}
    \item Because $\mat{e}_\mathrm{sup}$ is a fixed dimension vector, it requires a fixed number of enrolled speakers $N$.
    \item It is not permutation invariant. In other words, changing the order of two enrolled speaker embeddings will result in a different super embedding. Thus in the training process, each input needs to be transformed to all the $N!$ permutations to make the model robust. This unnecessarily makes the learning problem much harder.
\end{enumerate}

\subsubsection{ScorerNet}

In Eq.~\ref{eq:scorer_net}, we use a dedicated small feedforward neural network \emph{ScorerNet} to compute the attention scores. We prove that a simple cosine similarity (or dot-product) attention mechanism cannot be used for this problem. 

\begin{proof}
Assume we can use a simple cosine similarity attention mechanism:
\begin{equation}
    s_i^{(t)} = \cos( \mat{k}^{(t)} , \mat{e}_i ) .
\end{equation}
Assume that the input contains speech from the target speaker $A$ and an interference speaker $B$. Let $\mat{e}_A$ and $\mat{e}_B$ be the speaker embeddings of speaker $A$ and speaker $B$, respectively.

In an effectively working model, because the input contains speech from the target speaker $A$, the attentive embedding must be close to the embedding of speaker $A$:
\begin{equation}
    \mat{e}_\mathrm{att}^{(t)} \approx \mat{e}_A,
\end{equation}
thus
\begin{equation}
    \label{eq:cos_A}
    \cos( \mat{k}^{(t)} , \mat{e}_A ) \approx 1 .
\end{equation}
Because the input also contains speech from speaker $B$, according to symmetry, we must also have:
\begin{equation}
    \label{eq:cos_B}
    \cos( \mat{k}^{(t)} , \mat{e}_B ) \approx 1 .
\end{equation}
From Eq.~\ref{eq:cos_A} and Eq.~\ref{eq:cos_B}, we can easily infer that:
\begin{equation}
    \label{eq:cos_AB}
    \cos( \mat{e}_A , \mat{e}_B ) \approx 1 .
\end{equation}

However, because speaker $A$ and speaker $B$ are random different speakers, Eq.~\ref{eq:cos_AB} apparently does not hold. Thus, the hypothesis that a simple cosine similarity attention mechanism~\cite{kanda2020joint} can be used is incorrect.
\end{proof}

\subsubsection{Practical considerations in TFLite implementation}
Although the multi-user VoiceFilter-Lite model supports an arbitrary number of enrolled speaker embeddings as side input, when implementing the model in TFLite, there are additional constraints. As TFLite currently does not support inputs with an unknown dimension, we have to pre-define a maximal number of enrolled speaker embeddings, \emph{i.e.} $N$ in our implementation.

At runtime, when the actual number of enrolled speakers is smaller than $N$, we use \textbf{all-zero vectors} as the embeddings of the missing speakers. At training time, we also randomly replace some actual speaker embeddings by all-zero vectors to simulate this scenario. As all valid speaker embeddings are $\L_{2}$-normalized unit-length vectors, an all-zero vector will not conflict with any actual speaker in the embedding space.

\begin{table*}[!ht]
\centering
  \caption{Word Error Rate (\%) of streaming ASR system with different VoiceFilter-Lite (VFL) models and different number of enrolled users. The ASR model is trained on multi-domain data, and evaluated on vendor-collected speech queries.}
  \vspace{0.1cm}
  \label{tab:wer}
  \begin{tabular}{| c | c | c | c | c | c | c | c |}
    \hline
    \bf \multirow{2}{*}{Model} & \bf Enrolled   & \bf \multirow{2}{*}{Clean} & \multicolumn{2}{|c|}{\bf Non-speech noise (1$\sim$10 dB)} & \multicolumn{2}{|c|}{\bf Speech noise (1$\sim$10 dB)} & \bf \multirow{2}{*}{Size}\\ \cline{4-7}
    & \bf users & & \bf Additive & \bf Reverb & \bf Additive & \bf Reverb & \\ \hline
    \multicolumn{2}{|c|}{No VFL} & 15.3 & 21.1 & 29.1 & 56.0 & 54.5 & N/A \\ \hline
    Single-user VFL & 1 & 15.3 & 21.1 & 29.0 & 31.3 & 37.5 & 2.7 MB \\ \hline
    \multirow{4}{*}{Multi-user VFL} & 1 & 15.3 & 21.1 & 29.0 & 37.5 & 42.0 & \multirow{4}{*}{3.3 MB} \\ \cline{2-7}
    & 2 & 15.4 & 21.2 & 29.3 & 39.4 & 43.0 & \\ \cline{2-7}
    & 3 & 15.5 & 21.3 & 29.1 & 40.9 & 44.7 & \\ \cline{2-7}
    & 4 & 15.5 & 21.3 & 28.8 & 41.5 & 44.9 & \\ \hline
  \end{tabular}
\end{table*}

\section{System}
\label{sec:system}

\subsection{Personalized keyphrase detection}

We use VoiceFilter-Lite as part of our streaming personalized keyphrase detection system, which is illustrated in Fig.~\ref{fig:digram}. This system can be easily customized to accurately detect any phrase composed of words from a large vocabulary.
The system is implemented  with an  end-to-end  trained  ASR model~\cite{he2019streaming} and a text-independent speaker verification  model~\cite{wan2018generalized}. To  address  the  challenge  of  detecting  keyphrases  under  various  noisy  conditions, VoiceFilter-Lite is used as a speaker separation module in the feature frontend of the speaker verification model. In addition, an adaptive noise cancellation (ANC) algorithm~\cite{widrow1975adaptive,huang2019hotword} is included in the feature frontend of both ASR and speaker separation, which exploit cross-microphone noise coherence to reduce background noise. For further details, we refer readers to \cite{rikhye2021personalized}.

\subsection{Feature frontend}
In our experiments, a shared feature frontend is used for all speech models including speech recognition, speaker recognition, and VoiceFilter-Lite. This frontend first applies automatic gain control~\cite{prabhavalkar2015automatic} to the input audio, then extracts 32ms-long Hanning-windowed frames with a step of 10ms. For each frame, 128-dimensional log Mel-filterbank energies (LFBE) are computed in the range between 125Hz and 7500Hz. These filterbank energies are then stacked by 4 frames and subsampled by 3 frames, resulting in final features of 512 dimensions with a frame rate of 30ms.

\subsection{Speaker recognition}
\label{sec:spk_reg}
The speaker embeddings, \emph{a.k.a.} d-vectors, are computed using a text-independent speaker recognition model trained with the generalized end-to-end extended-set softmax loss~\cite{wan2018generalized,pelecanos2021dr}. Most of our training data are from a vendor collected multi-language speech query dataset covering 37 locales. We also added public datasets including LibriVox,
% VoxCeleb~\cite{nagrani2017voxceleb,chung2018voxceleb2},
CN-Celeb~\cite{fan2020cn},
TIMIT~\cite{garofolo1993darpa},
% Spoken Wikipedia Corpora~\cite{baumann2019spoken},
% BookTubeSpeech~\cite{pham2020toward}
and VCTK~\cite{yamagishi2019cstr}
to the training data for domain robustness. Multi-style training (MTR)~\cite{lippmann1987multi,ko2017study,kim2017generation} with SNR ranging from 3dB to 15dB is applied during the training process for noise robustness. The speaker recognition model has 3 LSTM layers each with 768 nodes and a projection size of 256. The output of the last LSTM layer is then linearly transformed to the final 256-dimension d-vector. 

\subsection{VoiceFilter-Lite}
The VoiceFilter-Lite model has 3 LSTM layers, each with 256 nodes, and a final fully connected layer with sigmoid activation. The noise type prediction network 2 LSTM layers, each with 128 nodes, and a fully connected layer with 64 nodes. In the multi-user setup, the PreNet has 3 LSTM layers, each with 128 nodes; the ScorerNet has two feedforward layers, each with 128 nodes. The new layers in PreNet and ScorerNet increased the quantized~\cite{alvarez2016efficient,shangguan2019optimizing} model size from the original 2.7MB to 3.3MB.

The training data of the VoiceFilter-Lite model consist of: (1) The LibriSpeech training set~\cite{panayotov2015librispeech}; and (2) a vendor-collected dataset of realistic English speech queries. For more details on VoiceFilter-Lite, please refer to ~\cite{Wang2020} and ~\cite{rikhye2021personalized}.

\section{Experiments}
\label{sec:exp}

\subsection{Multi-talker speech recognition}
\label{sec:exp_asr}
Our first group of experiments focuses on using VoiceFilter-Lite to improve streaming on-device ASR, especially for overlapped speech. Similar evaluations had been done in~\cite{Wang2020} for single-user cases. We use Word Error Rate (WER) as our metrics of ASR.

\subsubsection{Datasets}
\label{sec:wer_data}

For easier comparisons with previous work, instead of using the ASR model from the keyphrase detection system, here we use the same ASR model as the one described in~\cite{Wang2020}. This is an on-device streaming RNN-transducer model~\cite{graves2012sequence,he2019streaming} trained with data from multiple domains including YouTube and anonymized voice search. We evaluate this ASR model on a vendor-collected dataset of realistic speech queries, which consists of about 20,000 utterances from 230 speakers.

During evaluation, we apply different noise sources and room configurations to the data. We use ``Clean" to denote the original non-noisified data, although they could be quite noisy already.
The non-speech noise source consists of ambient noises recorded in cafes, vehicles, and quiet environments, as well as audio clips of music and sound effects downloaded from Getty Images\footnote{\url{https://www.gettyimages.com/about-music}}.
The speech noise source is a distinct development set without overlapping speakers from the testing set. We evaluate with two types of room conditions:  ``Additive" means directly adding the noise waveform to the clean waveform; ``Reverb" consists of 3 million convolutional room impulse responses generated by a room simulator~\cite{kim2017generation}. Each noise source was then applied to each utterance with an SNR that was drawn from a uniform distribution from $1$dB to $10$dB for both additive and reverberant conditions.

\subsubsection{Results}
\label{sec:asr_results}
The evaluation results are shown in Table~\ref{tab:wer}. From the table, we can see that multi-user VoiceFilter-Lite has similar performance as single-user VoiceFilter-Lite: under clean and non-speech noise conditions, there is no degradation in ASR performance; whereas under speech noise conditions, WER is significantly reduced. Unlike single-user VoiceFilter-Lite, multi-user VoiceFilter-Lite is able to work when there are multiple enrolled users. However, we observed that increasing the number of enrolled users decreased the WER improvement under speech noise conditions. This is because selecting the correct enrolled speaker in the presence of speech background noise is a hard problem.

\subsubsection{Other observations}

The PreNet and ScorerNet modules in Fig.~\ref{fig:attentive_embedding} can be viewed as a student network that attempts to distill some knowledge~\cite{hinton2015distilling} from the speaker recognition network described in Section~\ref{sec:spk_reg}. However, VoiceFilter-Lite applies to each frame of the input, but it can be very difficult to recognize the correct speaker on early frames of the input. During the training process, we observed that the values of most of the attention weights $\alpha_i^{(t)}$ are close to $0$, with just one close to $1$, suggesting that the attention mechanism learns to correctly select one out of the $N$ speaker profiles.

\subsection{Multi-talker speaker verification}
\label{sec:exp_mt_sv}

\begin{table*}[!ht]
\centering
  \caption{Equal Error Rate (\%) of our speaker verification system with different VoiceFilter-Lite (VFL) models and different number of enrolled users. Note that the number of users in this table is only for VFL, not for speaker verification. Speaker verification uses a predefined list of enrollment utterances.}
  \vspace{0.1cm}
  \label{tab:eer}
  \begin{tabular}{| c | c | c | c | c | c | c | c | c |}
    \hline
     & \bf \multirow{3}{*}{Room} &  & \multicolumn{6}{c|}{\bf EER (\%)} \\ \cline{4-9}
    \bf Noise  & & \bf SNR & \bf No  & \bf Single-user VFL & \multicolumn{4}{c|}{\bf Multi-user VFL} \\ \cline{5-9}
    \bf source &  &  \bf (dB) & \bf VFL & \bf 1 user & \bf 1 user & \bf 2 users & \bf 3 users & \bf 4 users \\ \hline
    \multicolumn{3}{|c|}{Clean} & 0.71 & 0.71 & 0.71 & 0.72 & 0.72 & 0.72 \\ \hline
    \multirow{6}{*}{Non-speech} & \multirow{3}{*}{Additive} & -5 & 5.05 & 4.99 & 5.03 & 5.04 & 5.05 & 5.05 \\ \cline{3-9}
    & & 0 & 2.20 &2.19 &2.21 &2.21 &2.22 &2.22
 \\ \cline{3-9}
    & & 5 & 1.47 &1.47 &1.49 &1.49 &1.50 &1.49
 \\ \cline{2-9}
     & \multirow{3}{*}{Reverb} & -5 & 7.76&7.65 &7.65 &7.76 &7.75 &7.78
 \\ \cline{3-9}
    & & 0 & 3.66&3.64 &3.70 &3.64 &3.65 &3.69
 \\ \cline{3-9}
    & & 5 & 2.09 &2.03 &2.06 &2.06 &2.12 &2.07
 \\ \hline
    \multirow{6}{*}{Speech} & \multirow{3}{*}{Additive} & -5 & 12.27 &4.04 &7.32 &9.34 &10.36 &10.99
\\ \cline{3-9}
    & & 0 & 8.21 &2.45 &3.90 &5.18 &5.73 &6.10
\\ \cline{3-9}
    & & 5 & 5.15 &1.68 &2.19 &2.78 &3.01 &3.14
\\ \cline{2-9}
     & \multirow{3}{*}{Reverb} & -5 & 17.08 &6.28 &10.76 &13.03 &14.11 &14.60
\\ \cline{3-9}
    & & 0 & 10.87 &3.50 &5.77 &7.15 &7.87 &8.19
\\ \cline{3-9}
    & & 5 & 6.42 &2.18 &3.19 &3.81 &4.17 &4.37\\ \hline
  \end{tabular}
\end{table*}

Our second group of experiments focuses on addressing the multi-talker speaker verification challenge, especially for the \emph{multi-user multi-talker} case. For example, when both speaker $A$ and speaker $B$ enroll their voices on the device (multi-user), and at runtime, speaker $A$ and speaker $C$ speaks at the same time (multi-talker), we expect the speaker verification system to \underline{accept} the input because it contains speech from one of the enrolled users (speaker $A$).

We evaluate the standard speaker verification task under various noise conditions with: (1) no VoiceFilter-Lite model; (2) single-user VoiceFilter-Lite model; (3) multi-user VoiceFilter-Lite model for single enrolled user; (4) multi-user VoiceFilter-Lite model for multiple enrolled users. Similar to the experiments in Section~\ref{sec:exp_asr}, the noise source can be either non-speech noise or an interfering speaker, and the room condition can be either additive or reverberant to simulate both near-field and far-field devices. The speaker verification model being evaluated is the same model which we used to generate the d-vectors for VoiceFilter-Lite, as described in Section~\ref{sec:spk_reg}.

\subsubsection{Datasets}
For this speaker verification evaluation, we use a vendor-provided English speech query dataset. The enrollment list comprises 8,069 utterances from 1,434 speakers, while the test list comprises 194,890 utterances from 1,241 speakers. The interfering speech are from a separate English dev-set consisting of 220,092 utterances from 958 speakers. The non-speech noise and room impulse responses are the same as described in Section~\ref{sec:wer_data}.

\subsubsection{Results}

The evaluation results are shown in Table~\ref{tab:eer}. From the table, we see that for clean and non-speech noise conditions, the EER of the speaker verification system does not change much when we apply single-user or multi-user VoiceFilter-Lite in the feature frontend.

When there is speech noise, single-user VoiceFilter-Lite provides the biggest improvement. The improvement of multi-user VoiceFilter-Lite is smaller than single-user model, but still quite significant compared to the no VoiceFilter-Lite case. Similar to what we previously observed in Section~\ref{sec:asr_results}, when the number of enrolled users increases, the performance of multi-user VoiceFilter-Lite degrades as the problem becomes harder.

\subsection{Keyphrase detection in the presence of ambient noise}
\label{sec:exp_keyphrase}

\begin{table*}[t]
\centering
\caption{End-to-end performance of our keyphrase detection on a variety of datasets with single- and multi-user VoiceFilter-Lite (VFL) models. Only the reverberant room condition is considered. The number of users are the number of enrolled speakers. SV, speaker verification.}
\vspace{0.1cm}
\label{tab:fr}
\centering
\begin{tabular}{|c|c|c|c|c|c|c|c|c|c|}
\hline
 \multirow{2}{*}{\textbf{\begin{tabular}[c]{@{}c@{}}Noise\\ Source\end{tabular}}} & \multirow{2}{*}{\textbf{\begin{tabular}[c]{@{}c@{}}SNR\\ (dB)\end{tabular}}} & \multirow{2}{*}{\textbf{\begin{tabular}[c]{@{}c@{}}No \\ SV \end{tabular}}} & \multicolumn{2}{c|}{\textbf{No VFL}} & \textbf{Single-user VFL} & \multicolumn{4}{c|}{\textbf{Multi-user VFL}} \\ \cline{4-10} 
 &  &  & \textbf{1 user} & \multicolumn{1}{l|}{\textbf{4 users}} & \textbf{1 user} & \textbf{1 user} & \textbf{2 users} & \textbf{3 users} & \textbf{4 users} \\ \hline
\multicolumn{10}{|c|}{\multirow{2}{*}{\textbf{\begin{tabular}[c]{@{}c@{}}Vendor-provided keyphrase queries\\ False Rejection Rate (\%)\end{tabular}}}} \\
\multicolumn{10}{|c|}{} \\ \hline
\multicolumn{2}{|c|}{Clean} & 4.23 & 4.23 & 4.36 & 4.23 & 4.30 & 4.30 & 4.30 & 4.30 \\ \hline
\multirow{3}{*}{Non-speech} & -5 & 21.8 & 23.6 & 24.6 & 23.6 & 24.1 & 24.5 & 24.5 & 24.5 \\ \cline{2-10} 
 & 0 & 7.62 & 8.16 & 8.58 & 8.16 & 8.22 & 8.25 & 8.25 & 8.25 \\ \cline{2-10} 
 & 5 & 3.63 & 3.72 & 3.90 & 3.72 & 3.75 & 3.84 & 3.84 & 3.84 \\ \hline
\multirow{3}{*}{Speech} & -5 & 82.1 & 93.7 & 94.6 & 82.1 & 84.6 & 85.1 & 88.4 & 89.4 \\ \cline{2-10} 
 & 0 & 60.8 & 78.5 & 79.5 & 60.8 & 62.1 & 65.8 & 67.4 & 69.8 \\ \cline{2-10} 
 & 5 & 22.6 & 28.4 & 29.5 & 22.6 & 23.1 & 24.1 & 25.2 & 25.2 \\ \hline
\multicolumn{10}{|c|}{\multirow{2}{*}{\textbf{\begin{tabular}[c]{@{}c@{}}YouTube (no queries)\\ False Acceptance / hour\end{tabular}}}} \\
\multicolumn{10}{|c|}{} \\ \hline
\multicolumn{2}{|c|}{} & 0.275 & 0.00985 & 0.0346 & 0.00985 & 0.00985 & 0.0197 & 0.0346 & 0.0346 \\ \hline
\end{tabular}
\end{table*}

We previously demonstrated that one application of speaker separation using VoiceFilter-Lite is to reduce false rejects in a personalized keyphrase detection system when the background contains speech noise ~\cite{rikhye2021personalized}. Specifically, by placing VoiceFilter-Lite in the feature frontend of speaker verification, we can mitigate speaker identification errors due to interfering speech, and as a result reduce false rejections. Thus, in our last group of experiments, we evaluate the overall performance of the personalized keyphrase detection system with multiple enrolled speakers under various background noise conditions. Specifically, we evaluate based on the following two metrics: (1) the number of false acceptance per hour (FA/h), which measures how many keyphrases are incorrectly accepted by the system; and (2) the false rejection rate (FRR), which measures the percentage of true keyphrases that are ignored by the system.

\subsubsection{Datasets}

To evaluate FA/h, we used a dataset consisting of 156 hours of English speech from curated and hand-annotated YouTube videos~\cite{narayanan2019longform}. This dataset is designed to mimic television/radio noise, and contains no true keyphrases. As such, phrases in this dataset that trigger the detection system are considered false accepts.

To evaluate FRR, we used a set of vendor-provided keyphrases consisting of 61,555 utterances from 250 speakers with an average of 240 utterances per speaker. This dataset contained commonly used keyphrases such as ``remind me to set an alarm", ``turn off the lights", and ``set a timer". 

For both evaluations, we first selected four speakers at random from the vendor-provided dataset. Next, for each speaker, we generated d-vectors from four standard enrollment utterances (\emph{e.g.} ``Hey Google, remind me to water my plants"). The number of speakers enrolled however varied depending on the model. For example, to test a 4-user VoiceFilter-Lite, we enrolled all four speakers, while to test the 1-user VoiceFilter-Lite model, we enrolled just a single speaker and appended three all-zero vectors to that embedding. FRR evaluation was performed on utterances from all four speakers, such that all models were evaluated on the same number of utterances. We repeated this process for all 4-speaker combinations in our dataset. In order to mimic ambient conditions, we augmented the dataset with either speech or non-speech background noise with reverberation at three different SNR levels using MTR. Therefore the speech background noise condition contains overlapping speech from both the enrolled and non-enrolled speakers. The ``Clean'' condition in Table~\ref{tab:fr} refers to data that is not augmented. 

\subsubsection{Results}

The overall end-to-end performance of our keyphrase detection system is shown in Table~\ref{tab:fr}. We observed that including speaker verification with either a single-user or a multi-user VoiceFilter-Lite model with one enrolled user significantly decreased FA/h from 0.275 to 0.00985 (rel. \textbf{96.4\%}). With more than one enrolled speaker, there is a greater likelihood of falsely identifying the speaker on the YouTube negative dataset, leading to a marginal increase in the FA/h to 0.0346 in the 4-user case. Despite this increase, speaker verification with VoiceFilter-Lite in its feature frontend is sufficient to significantly reduce false triggering. 

As we have demonstrated previously~\cite{rikhye2021personalized}, adding speaker verification to the keyphrase detection system increased FRR by 30.8\% in the multi-talker and 7.09\% in the non-speech case (SNR = 0dB). This was primarily due to incorrect speaker verification, rather than incorrect speech recognition. This increase in FRR was larger when four speakers were enrolled. Adding a single-user VoiceFilter-Lite model to the feature frontend of speaker verification reduced the FRR from 78.5\% to 60.8\%, resulting in a \textbf{22.5\%}  reduction in FRR in the SNR = 0dB multi-talker case relative to the model with only speaker verification. Similarly, when we enroll only one user, the multi-user VoiceFilter-Lite model reduces FRR to 62.1\% (rel.\textbf{ 20.1\%}). This suggests that the single-user and a multi-user VoiceFilter-Lite model with a single enrolled user perform similarly under these conditions. Furthermore, we observe that both the single-user and multi-user VoiceFilter-Lite models perform similarly in the presence of non-speech background noise.

Interestingly, as we enroll more users with the multi-user VoiceFilter-Lite model, we notice an increase in FRR relative to the single-user VoiceFilter-Lite model for both speech and non-speech noise conditions. This is because the attention mechanism now has to select between multiple speakers in the presence of background noise, which is a fundamentally difficult problem especially if the speakers have similar vocal patterns. However, despite this degradation in performance, the multi-user VoiceFilter-Lite with 4 enrolled users has a lower FRR (rel. \textbf{12.2\%}, SNR = 0dB speech case) than the 4-user system without VoiceFilter-Lite. 

Altogether, we demonstrate that by using an attention mechanism to select between $N$ user embeddings, we can extend the benefit of having a VoiceFilter-Lite speaker separation model in the speaker verification feature frontend to multiple enrolled users. This system will greatly reduce both false triggering and false rejections in the presence of ambient noise.

\section{Conclusions}
\label{sec:conclusions}

In this paper, we describe a generic solution that allows speaker-conditioned speech models to support scenarios where multiple users have enrolled their voices. This is a very common use case for shared devices, such as smart home speakers. The core idea is to build a permutation-invariant attentive speaker embedding from all enrolled speaker embeddings. We implemented this idea for VoiceFilter-Lite, a streaming target voice separation model for on-device ASR and speaker verification. Experiments show that our multi-user VoiceFilter-Lite model is able to significantly reduce multi-talker speech recognition and speaker verification errors with up to four enrolled users. At the same time, the improvement becomes less when the number of enrolled users increases, as the problem also becomes more difficult.

One future research direction would be to explore alternative implementations of the attention mechanism to ensure that the correct target speaker profile is selected on each trial and to better match the multi-user VoiceFilter-Lite performance with single-user models. Another interesting future direction would to extend this multi-user VoiceFilter-Lite model to separate speech from two enrolled users if they talk simultaneously. Besides this, the same multi-user solution for VoiceFilter-Lite can be easily used for other speaker-conditioned speech models, such as personal VAD and personalized ASR.

% References should be produced using the bibtex program from suitable
% BiBTeX files (here: strings, refs, manuals). The IEEEbib.bst bibliography
% style file from IEEE produces unsorted bibliography list.
% -------------------------------------------------------------------------
% \clearpage
% \pagebreak
\bibliographystyle{IEEEbib}
\bibliography{refs}

\end{document}